\documentclass[doublecol,a4paper]{epl2}
\usepackage{color}
\usepackage{amsmath}
\usepackage{amssymb}
\usepackage{enumerate}
\usepackage[colorlinks=true, pdfstartview=FitV, linkcolor=blue, citecolor=red, urlcolor=black, breaklinks=true]{hyperref}
\usepackage{subfigure}
\newcommand{\be}{\begin{equation}}
\newcommand{\ee}{\end{equation}}
\newcommand{\ben}{\begin{eqnarray}}
\newcommand{\een}{\end{eqnarray}}
\newcommand{\bes}{\begin{subequations}}
\newcommand{\ees}{\end{subequations}}

\newcommand{\sech}{{\rm sech}}

\begin{document}

\title{First order formalism for thick branes in modified gravity with Lagrange multiplier}

\author{D. Bazeia\inst{1} \and D. A. Ferreira\inst{1}\and D. C. Moreira\inst{2}}

\shortauthor{D. Bazeia \etal}

\institute{
\inst{1} Departamento de F\'isica, Universidade Federal da Para\'iba, 58051-970, Jo\~ao Pessoa, PB, Brazil\\
\inst{2} Faculdade Uninassau Petrolina, 56308-210, Petrolina, PE, Brazil}

\abstract{ This work discuss the construction of braneworld solutions in modified gravity with Lagrange multipliers. We examine the general aspects of the model and present a first order formalism that help us to find analytic solutions of the equations of motion. We also investigate some explicit models, analyse linear stability of the metric and comment on how to relate models investigated in other works to the ones examined in the present study.}


\maketitle

{I. \it Introduction.} Modified gravity theories have played an interesting role in the discussion of some unsolved problems in High Energy Physics. In particular, there are several distinct lines of research in scalar-tensor models that provide interesting suggestions to the dark energy problem, the negative pressure fluid responsible for the current accelerated expansion of the Universe \cite{1,2}. A class of models which came to light in the early 1990s to address the problem of cosmological singularities \cite{3,4} has received renewed attention in recent years with the use of Lagrange multipliers to study dark energy \cite{5,6,7,8}. In this approach a multiplier is usually inserted into the fundamental action to play the role of an auxiliary field that sets a constraint on the system such that the norm of the scalar field gradient is nontrivially related to the field through a scalar potential. 

Modified gravity with Lagrange multiplier can also generate new cosmological solutions when associated to other modified theories, such as $F(R)$ models \cite{7,8,9}, Gauss-Bonnet gravity \cite{10,11,12} and covariant renormalisable gravity \cite{13,14}. Moreover, another subset of modified gravity with Lagrange
multiplier is known as {\it mimetic gravity} \cite{15}, which emerged from the idea of rewriting the background metric as a nontrivial coupling between a scalar field and another auxiliary metric. Later, however, it was observed that this is equivalent to the addition of a constraint in the action of the model \cite{16,17}. Although originally requiring only a factorization of the metric, which {\it a priori} should not change the covariant form of the equations, the field equations in mimetic gravity are different from the Einstein equation even in the absence of matter. In this sense, exploring new possibilities around these setups can be very instructive to understand the behavior of gravitational solutions and look for other properties not yet reached in other gravitational models.

Another important class of gravitational solutions addressed in modified gravity theories are the braneworlds, which have been extensively studied in the last twenty years \cite{18,19,20,21,22,23,24,25,26,27,28,29,30,31,32},
motivated in particular to understand the mass hierarchy problem, but only recently have been explored in modified gravity with Lagrange multipliers \cite{33,34}. In these models the scalar field acts as a source of gravity, which generates a brane in a five-dimensional bulk consisting of a four-dimensional flat spacetime with Poincar\'e symmetry and a single extra spatial dimension of infinite extent. Gravity is localized in the brane when there is a normalisable graviton zero mode related to the four-dimensional  invariance. In this work we present a first order formalism for the construction of branes in modified gravity with Lagrange multiplier. Such formalism already exists in other contexts \cite{35,36,37}, but doing so here may lead to further discussions related to other inherent peculiarities of branes arising when Lagrange multipliers play a role. In the models to be studied here, the degrees of freedom are associated to the source scalar field and the warp factor that controls the geometry. The constraint imposed by the Lagrange multiplier makes the scalar field kinetic term to be determined by a potential that depends only on the source scalar field. Thus, we can choose such potential in order to reduce the scalar field equation of motion to a first order differential equation that does not couple to the other degree of freedom associated to the warp factor. In this sense, the scalar field feeds the geometry, and we then propose another first order equation to describe the warp function to set up the conditions for the first order equations to control the equations of motion of the modified gravity model.  

Here we take advantage of the freedom engendered by the presence of two potentials, one related to the Lagrange multiplier and the other to be adjusted appropriately, to find solutions capable of inducing gravity localization. In these systems, the scalar field acts independently but can generate branes with well-behaved geometry. We illustrate the main results examining models of current interest, in particular the case of brane in mimetic gravity in the sense of \cite{15}. Also, a linear stability analysis is implemented to show that the braneworld scenario is robust, since its gravitational sector is stable against small fluctuations of the metric. In this sense, assembling a first order formalism in these models follows the idea of finding a systematic way to obtain analytic brane solutions which can help in the search for new properties and predictions on gravitational scenarios.

The letter is organized as follows. In Sec. II we introduce the model, write the equations of motion and then implement the first order formalism and discuss some of its properties. In Sec. III we examine braneworld models focusing on the obtention of analytical solutions and in Sec. IV we study the stability of the models under small fluctuations of the metric. We end the investigation in Sec. V with some comments and discussions.

{II. \it Formalism.} In this work we consider thick brane models driven by scalar fields in $(4,1)$ spacetime dimensions for a class of modified theories of gravity with Lagrange multiplier. The basic model is described by the action
\be \label{action}
S=\!\int \! d^{5}x\sqrt{|g|}\!\left(\!-\frac{R}{4}\!+\!I\left(\frac{1}{2}\partial_{a}\phi\partial^{a}\phi+\!V(\phi)\!\right)\! - U(\phi)\!\right)\!.
\ee
Here, $g$ is the determinant of the background metric $g_{ab}$ ($a,b=0,1,2,3,4$), $R$ is the Ricci scalar, $I$ represents the Lagrange multiplier,
$\phi$ denotes the scalar field and $V(\phi)$ and $U(\phi)$ are two potentials. As we shall see later, these potentials are responsible for the scalar field solution and the modelling of the brane, respectively. Also, the scalar function $I$ acts as a Lagrange multiplier, and its equation of motion imposes a constraint on the scalar field $\phi$ to be specified along with the potential $V(\phi)$.

The variation of the action \eqref{action} in terms of the scalar field $\phi$ and the background metric $g_{ab}$ leads us to the field equations
\bes\label{fe}\ben\label{eom}
&&I\left(g^{ab}\nabla_{a}\nabla_{b}\phi-\dfrac{dV}{d\phi}\right)+\partial_{a}I\partial^{a}\phi+\dfrac{dU}{d\phi}=0,\\[3pt] \label{einstein}
&&G_{ab}-2T_{ab}=0.
\een\ees
The constraint provided by the Lagrange multiplier becomes 
\be \label{constrain}
\frac{1}{2}g^{ab}\partial_{a}\phi\partial_{b}\phi+V(\phi)=0\,,
\ee
and the energy-momentum tensor is given by
\begin{equation}\label{em}
T_{ab}=I\partial_{a}\phi\partial_{b}\phi+g_{ab}U(\phi).
\end{equation}
The constraint provided by the Eq. \eqref{constrain} was used to simplify $T_{ab}$, so we can interpret the Lagrange multiplier as a field source which effectively modifies the scalar field dynamics.

The background geometry has four-dimensional  invariance and a single extra dimension of infinite extent. It is given by the general {\it ansatz}
\be \label{warped}
ds^{2}=e^{2A}\eta_{\mu\nu} dx^{\mu}dx^{\nu}-dy^{2}\,,
\ee
where the coordinate $y$ denotes the extra dimension, $\eta_{\mu\nu}=\text{diag}(1,-1,-1,-1)$ is the four-dimensional Minkowski metric with
$\mu,\nu=0,1,2,3$ and $A=A(y) $ is the warp function. The metric \eqref{warped} is the most general ansatz one can choose which holds four-dimensional Poincar\'e  symmetry. For simplicity, we  also consider that the scalar field is static and depends only on the extra dimension. In this way, equations \eqref{eom} and \eqref{einstein} become
\bes\label{systeq1}\ben \label{fieldeq}
&&I\left(\phi''+4A'\phi'+\frac{dV}{d\phi}\right)+I'\phi'-\frac{dU}{d\phi}=0,\\[3pt]
\label{E3}
&&3A'^{2}+U(\phi)-I\phi'^{2}=0,\\[3pt]
\label{E2}
&&A''+\frac{2}{3}I\phi'^{2}=0,
\een\ees  
and the constraint \eqref{constrain} now reduces to
\begin{equation}
\label{constrain2}
\phi'^{2}=2 V(\phi),
\end{equation}
where the prime denotes derivative with respect to the extra spatial coordinate $y$. We have a system of four differential equations, but it is not completely independent since if we derive the equation \eqref{E3} once again with respect to the extra coordinate and then use equations \eqref{E2} and \eqref{constrain2}, we recover \eqref{fieldeq}. We can also find the equation \eqref{fieldeq} by using Bianchi identity. 

We now concentrate in the task of finding solutions to the above model. It has been recently used in \cite{34,35} to investigate braneworld scenarios, and there the authors analysed internal structures and gravitational resonances. Our main motivation here is to describe a systematic procedure to obtain solutions of these braneworld models using a first order formalism to be developed below. An extra gain within the first order formalism is the possibility to expand the discussion on the construction of new solutions emerging in the above context. In order to move forward towards the development of the formalism, we recall the necessity to include auxiliary functions that help us to solve the equations of motion in a simplified way. If we focus on Eq. \eqref{constrain2}, we can introduce the function $\omega=\omega(\phi)$ such that
\be \label{potV}
V(\phi)=\frac{1}{2}\left(\dfrac{d\omega}{d\phi}\right)^{2},
\ee
and this naturally leads us to the first order equation
\be\label{phifo}
\phi'=\dfrac{d\omega}{d\phi}.
\ee
Moreover, we also include another function $W=W(\phi)$ to control the warp function, such that
\be\label{warpfo}
A^\prime=-\frac23 W(\phi).
\ee
We then use the above equations into Eq. \eqref{systeq1} to get
\be\label{potU}
U(\phi)=\dfrac{d\omega}{d\phi}\dfrac{dW}{d\phi}-\frac{4}{3}\,W^{2},
\ee
which is imposed to ensure consistency with the equations \eqref{systeq1} and \eqref{constrain2}. In particular, the Lagrange multiplier in this case has to obey
\be\label{LM} 
I(\phi)\frac{d\omega}{d\phi}=\frac{dW}{d\phi}.
\ee 
This equation shows that the auxiliary functions $W(\phi)$ and $\omega(\phi)$, in addition to dictating how the scalar field selfinteracts and inducing the expression of the warp factor, also have an effect on the dynamics of the scalar field. 

Before investigating specific models, let us comment on some interesting issues. First, the choice of the function $\omega(\phi)$, and thus the potential $V(\phi)$, directly implies the choice of the scalar field $\phi$, which obeys the equation \eqref{phifo}. However, the choice for the function $W(\phi)$ cannot be made carelessly, because of the Lagrange multiplier \eqref{LM}. The function $W(\phi)$ has to be combined with the scalar field solution given by equation \eqref{phifo} to make the warp function that obey the equation \eqref{warpfo} to provide models with gravity localization. Second, since \eqref{phifo} and \eqref{warpfo} are first order equations, they open the possibility to build analytic thick brane models where the scalar field can present distinct behavior. When choosing how to combine the scalar field solution with the function $W(\phi)$, one should care about the building of regular warp factor, such that 
$\lim_{y\to\pm\infty} \exp({2A(y)})\to 0.$
In this sense, the scalar field profile and the nonlinearities encoded in the choice of the function $W(\phi)$ have to be conveniently adjusted to give rise to robust braneworld scenarios. We notice, for instance, that the $00$-component of the energy-momentum tensor, given by $T_{00}=\rho(y)=\exp({2A}) U$
can under the use of the first order equations be expressed as the total derivative 
\be 
\rho(y)=\frac{d}{dy}(We^{2A}),
\ee
so the total energy of the brane vanishes for any well-behaved function $W(\phi)$. Moreover, we can calculate the Kretschmann scalar; in the above model it is given by
\be
K=16 {A''}^2 + 32{{ A'}}^2{A''}+40 {A'}^4\,,
\ee  
and should be examined to see if the gravitational sector of the model behaves adequately, in terms of the extra spatial dimension.

We also notice that for the two potentials $V(\phi)$ and $U(\phi)$ that are given by equations \eqref{potV} and \eqref{potU}, the first order equations \eqref{phifo} and \eqref{warpfo} solve the equations of motion when the Lagrange multiplier obeys \eqref{LM}. Moreover, if we set the Lagrange multiplier to unity, the model \eqref{action} becomes the standard model with potential $U_S(\phi)=U(\phi)-V(\phi)$. In this case, the equation \eqref{LM} imposes that $d\omega/d\phi=dW/d\phi$, and equations \eqref{potV} and \eqref{potU} lead to 
\be  
U_S(\phi)=\frac12\left(\frac{dW}{d\phi}\right)^2-\frac43\; W^2,
\ee
which is the form required for the potential of the standard gravity braneworld model, capable of generating a thick brane in the presence of the source scalar field, governed by the first order equations \eqref{phifo} and \eqref{warpfo}. 

Another issue is that the choice $\omega(\phi)=\phi$ leads us to the solution $\phi(y)=y$. In this case, if we choose
\be  
W(\phi)=\frac{3{\kappa}}{2}\; \text{sign}(\phi),
\ee 
with $\kappa$ being a real constant, the warp function has the solution $A(y)=-\kappa\, |y|$, and the warp factor leads to the line element
\be \label{warped2}
ds^{2}=e^{-2\kappa|y|}\eta_{\mu\nu} dx^{\mu}dx^{\nu}-dy^{2}\,.
\ee
Thus, we recover the solution presented in the Randall-Sundrum model \cite{18}, which describes the thin brane model. 

{III. \it Models.} Let us now discuss some models with distinct behavior, which can be addressed by using the first order formalism uncovered in the previous Section.

\begin{figure}[t!]
\begin{center}
\subfigure[$~$Warp factor]{\includegraphics[width=0.8\linewidth]{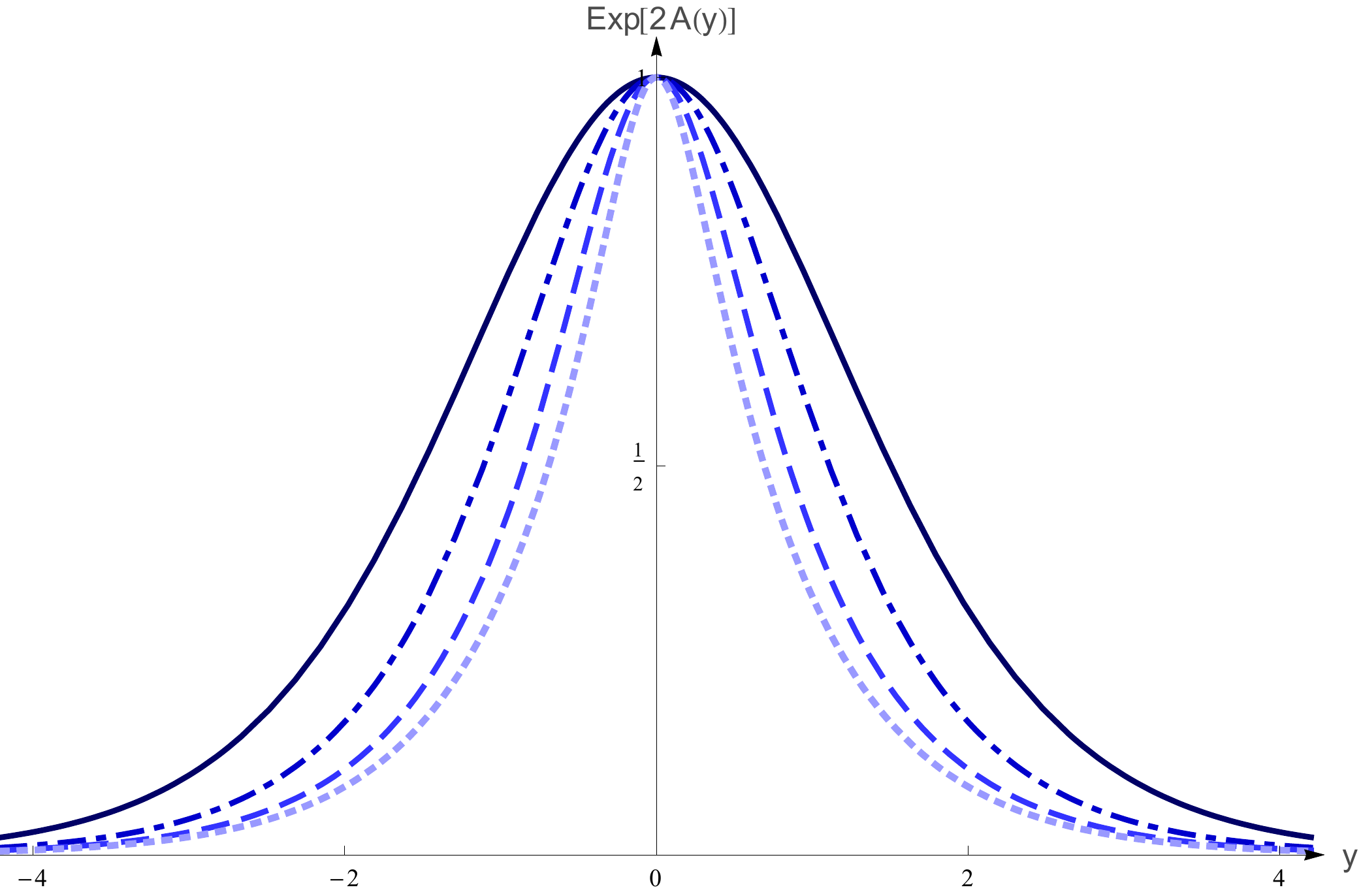}\label{fig1a}}
\subfigure[$~$ Energy density]{\includegraphics[width=0.8\linewidth]{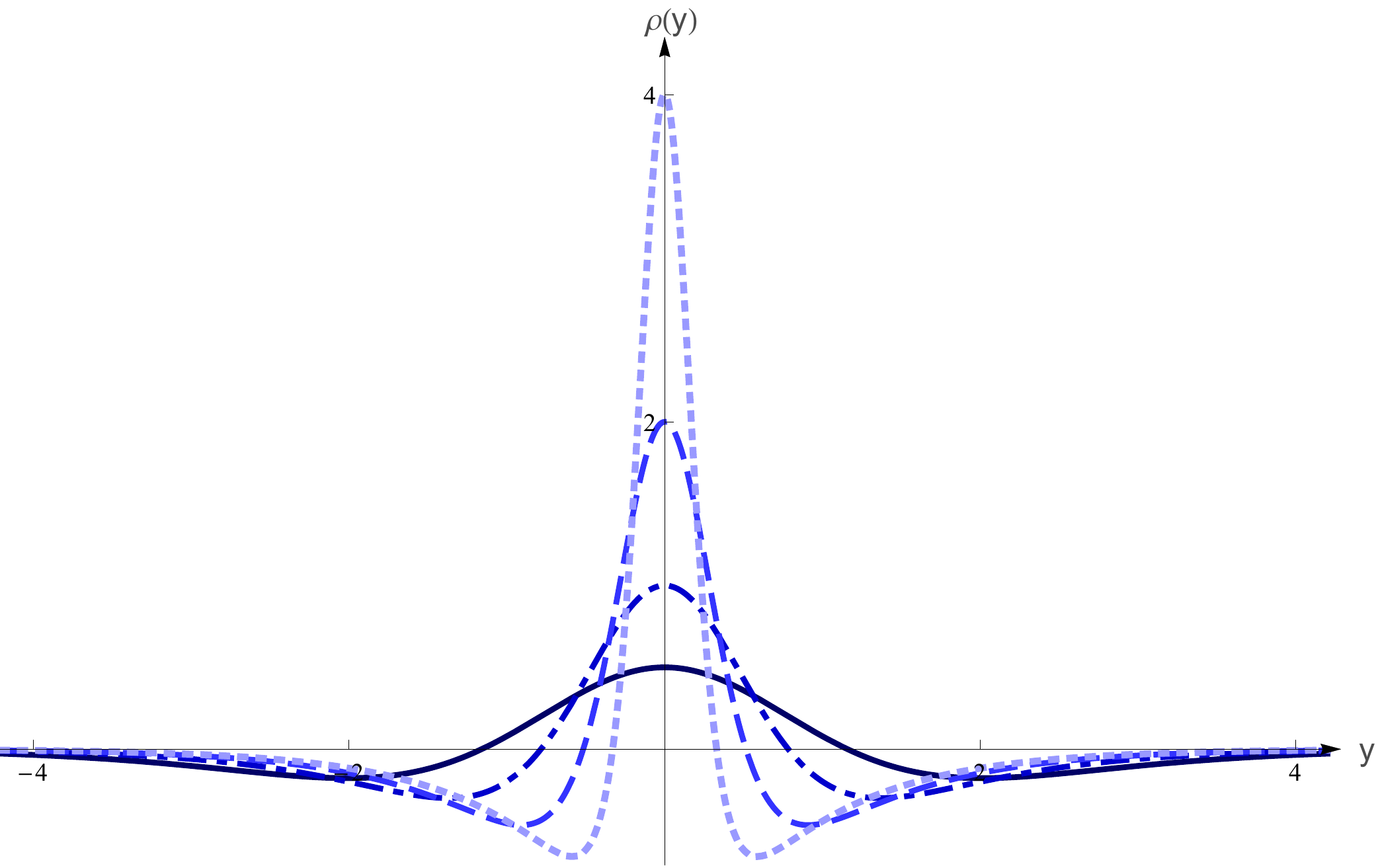}\label{fig1b}}
\end{center}
  \caption{First model. Warp factor (top panel) and energy density (bottom panel) for $a=1$. The solid (dark blue), dot-dashed (blue), dashed (light blue) and dotted (lighter blue) lines refer to $\alpha=1/2$, $1$, $2$ and $4$, respectively.}\label{fig0}
  \end{figure}

{III.1. \it First model.} Let us first consider the case with $W(\phi)=\phi$. Here we get
\be\label{warp0}
A'=-(2/3)\, \phi.
\ee
The scalar field in governed by the first order equation \eqref{phifo}, so we need $w(\phi)$ which is now proposed to be
\be  
\omega(\phi)=\alpha\, a\,\phi-\frac{\alpha}{3a}\,\phi^3,
\ee 
where $a$ and $\alpha$ are positive real parameters. In this case, the scalar field solution has the form
\be\label{phi0}
\phi(y)=a \,\tanh(\alpha\, y),
\ee
and $I(\phi)=a/(\alpha(a^2-\phi^2))$. We then have
\be  
e^{2A(y)}=\sech^{(4a/3\alpha)}(\alpha y),
\ee
and the energy density
\be  
\rho(y)\!=\!\alpha\, a\,\sech^{(4a/3\alpha)}\!\left(\!1\!-\!\left(\frac{4a}{3\alpha}+1\!\right)\!\tanh^2(\alpha y)\!\right).
\ee
The warp factor and energy density are depicted in Fig. \ref{fig0} for $a=1$ and for some values of $\alpha$.

We now recall the work \cite{22}, which unveiled an interesting way to reach the thin brane limit starting from a thick brane. The idea was to transform the smooth kinklike profile of the scalar field into an abrupt steplike configuration. It can also be implemented here, in the limit of a very high parameter $\alpha$. The point is that the kinklike solution \eqref{phi0} in the limit $\alpha\to\infty$ leads us to the solution $\phi(y)=a\, \text{sign}(y)$, such that the warp function becomes $A(y)=-(2a/3)|y|$. Thus, if one sets $a=3\kappa /2$ we get back to the thin  brane described in \cite{18} once again. 

We have checked that the Kretschmann scalar behaves nicely as a function of the extra spatial dimension, for a wide range of values of $\alpha$. We have also noticed that it diverges at $y=0$ in the limit $\alpha\to\infty$, and this is consistent with the divergence of the Kretschmann scalar for the thin brane described in \cite{18}. Interestingly, it was shown before that the thin brane can be obtained with $\omega(\phi)=\phi$, and an appropriate choice of $W(\phi)$. And above we have found that the thin brane can also be built in the case of $W(\phi)=\phi$, with an appropriate choice of $\omega(\phi)$.

{III.2. \it Second model.} The second model is described by the pair of functions
\begin{equation}\label{wW}
\omega(\phi)=\phi-\frac{\phi^{3}}{3};\;\;\;\;\;W(\phi)=\phi^{2n+1}\,,
\end{equation}
with $n=0,1,2,3,\cdots$. Here we choose to insert a parameter in the model only in the $W$ function, leaving the function $\omega$ with no parameters. It is straightforward to obtain the solution of the scalar field and the expression for the Lagrange multiplier from equations \eqref{phifo} and \eqref{LM}, where one finds $\phi(y)=\tanh(y)$
and $I(\phi)=(1+2n)({\phi^{2n}}/({1-\phi^2}))$. Since the function $\omega(\phi)$ has no parameters, the scalar field solution is fully determined, and represents the standard {\it kinklike} solution. The warp factor, on the other hand, will depend on $n$; the
Eq. \eqref{warpfo} in this case reads
\be\label{wf1}
A'=-\frac{2}{3}\tanh^{2n+1}(y)\,,
\ee
and has the solution
\be\label{wfs1}
A(y)=-\frac{\tanh^{2(n+1)}(y)}{3(n+1)}\,{}_{2}F_{1}\left(1,{n+1},{n+2},\tanh^{2}(y)\right),
\ee
where ${}_{2}F_{1}\left(\cdots\right)$ denotes hypergeometric function. Furthermore, the energy density associated to this model is given by
\be \label{ed1}
\rho(y)\!=\!e^{2A}\!\left(\!(2n\!+\!1)\, \sech^{2}(y)\tanh^{2n}(y)\!-\!\frac{4}{3}\tanh^{2(2n\!+\!1)}(y)\!\right).
\ee

The respective warp factor for the warp function \eqref{wfs1} is depicted in Fig. \ref{fig1a} for some values of $n$. The parameter $n$ controls the thickness of the brane and makes it smoother and flatter in the vicinity of the region $y = 0$, where the brane is located. The energy density \eqref{ed1} is depicted in Fig. \ref{fig1b}, where one can identify a splitting effect for $n\neq 0$, indicating that the parameter $n$ controls the internal structure of the brane. This behavior is similar to that presented by models investigated in \cite{40,41,43,44}, forcing the energy density to behave at $y=0$ as it behaves asymptotically. And yet, the Kretschmann scalar also behaves with no problem in the case.

\begin{figure}[t!]
\begin{center}
\subfigure[$~$Warp factor]{\includegraphics[width=0.8\linewidth]{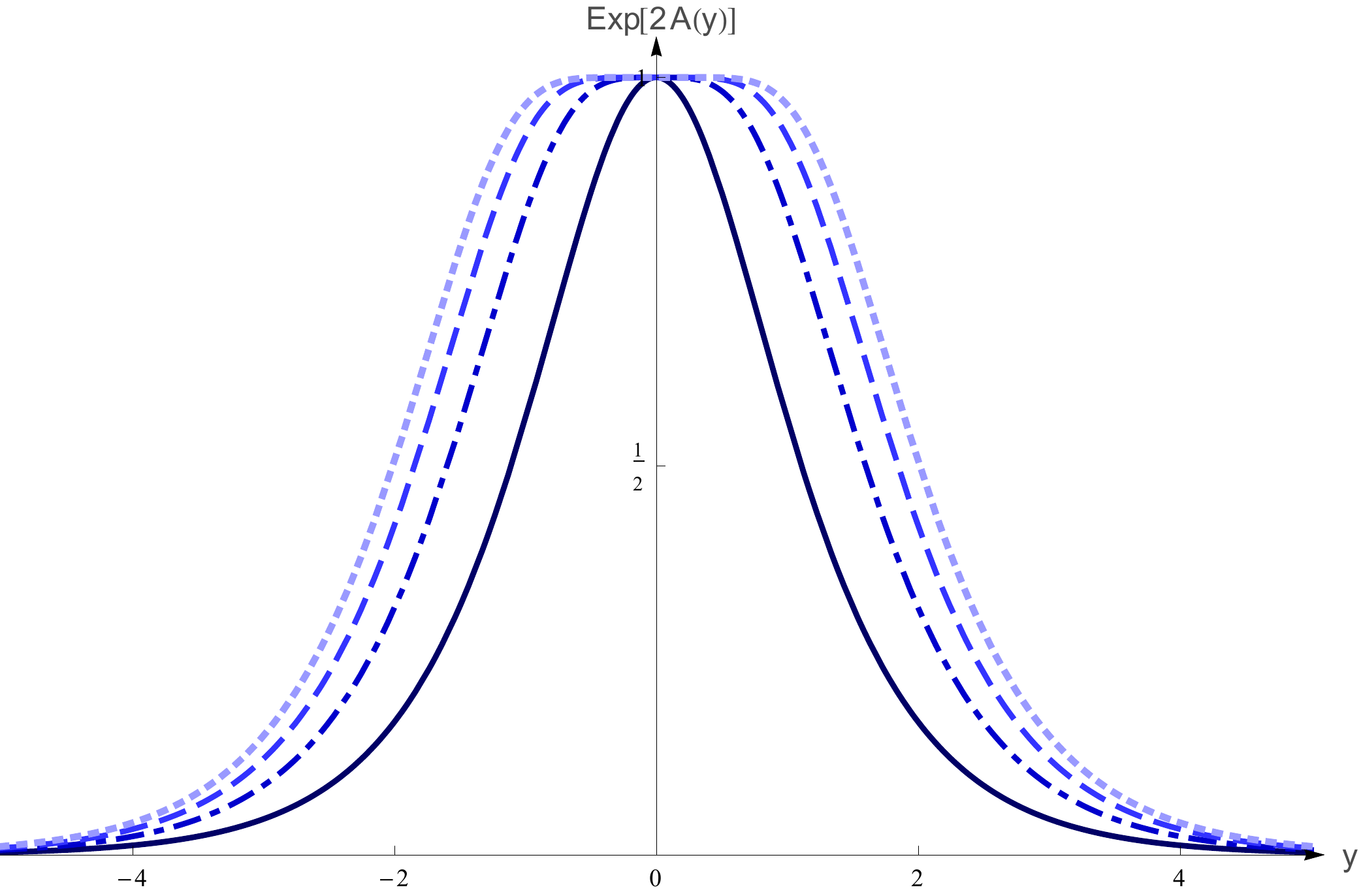}\label{fig1a}}
\subfigure[$~$ Energy density]{\includegraphics[width=0.8\linewidth]{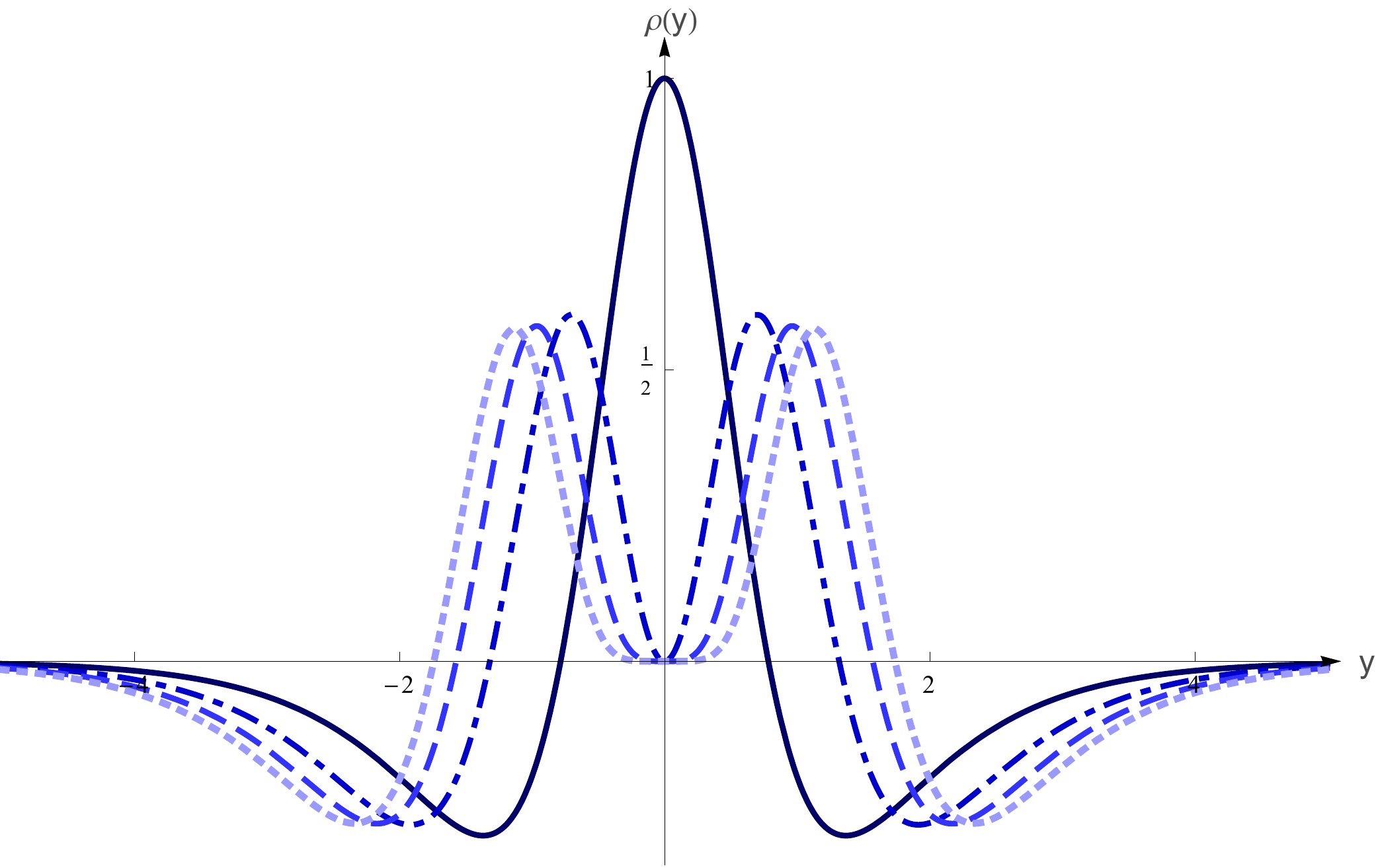}\label{fig1b}} 
\end{center}
\caption{Second model. Warp factor (top panel) and energy density (bottom panel). The solid (dark blue), dot-dashed (blue), dashed (light blue) and dotted (lighter blue) lines refer to $n=0$, $n=1$, $n=2$ and $n=3$, respectively.}\label{fig1}
  \end{figure}

\begin{figure}[t!]
\begin{center}
\subfigure[$~$ Warp factor]{\includegraphics[width=0.8\linewidth]{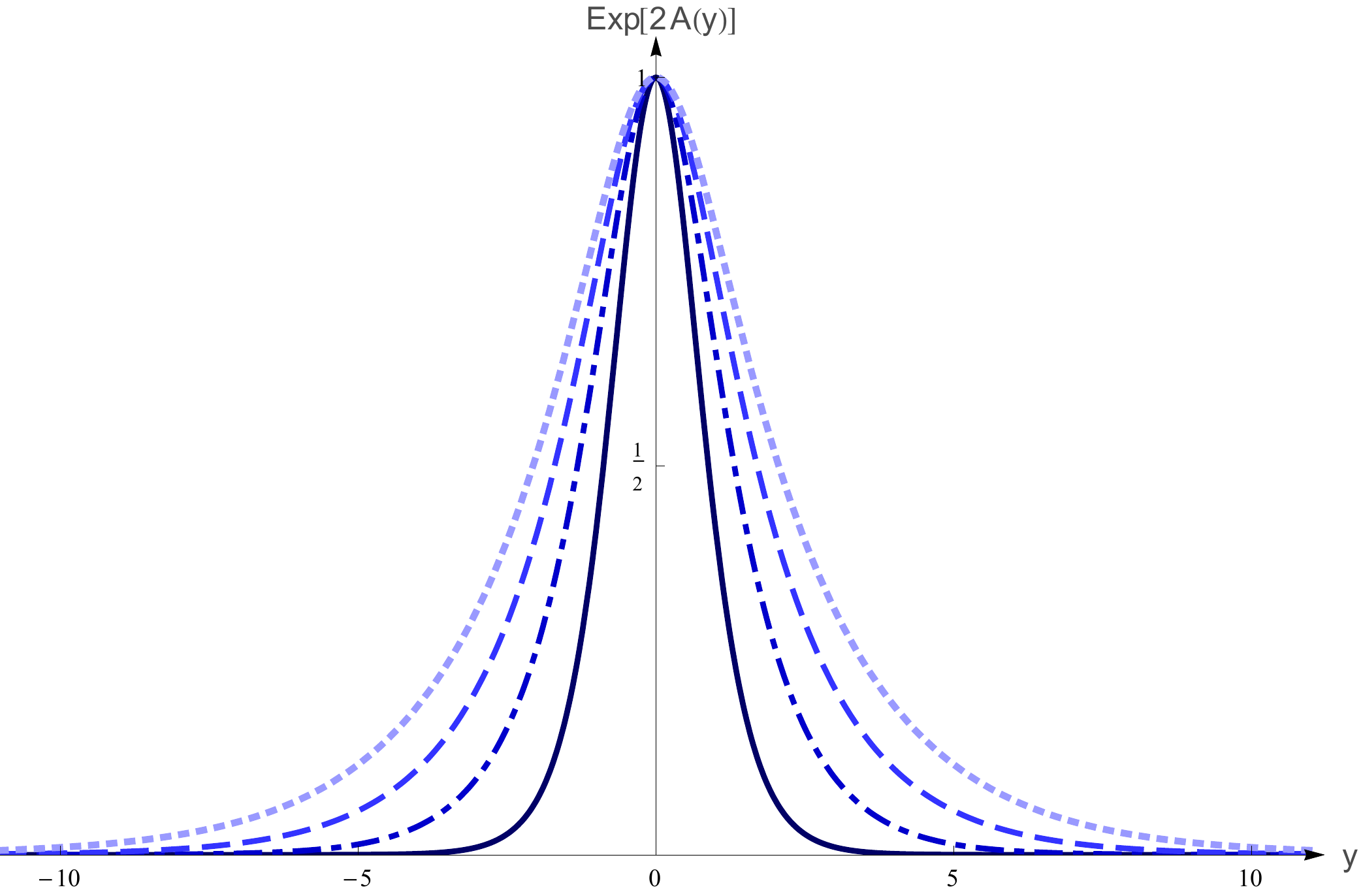}\label{fig3aa}} \hspace{5mm}
\subfigure[$~$ Energy density]{\includegraphics[width=0.8\linewidth]{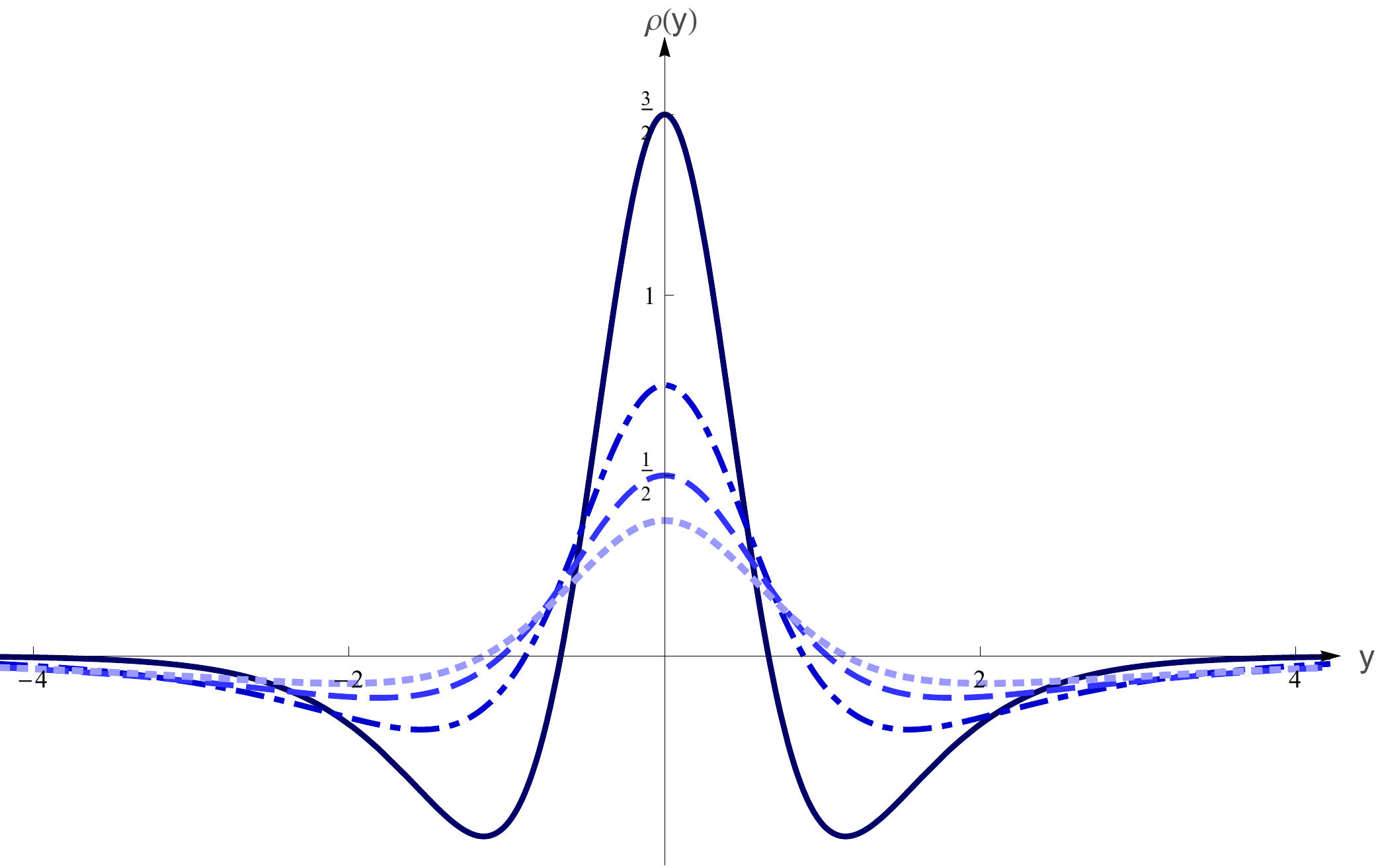}\label{fig3bb}} 
\end{center}
\caption{Third model. Warp factor (top panel) and energy density (bottom panel). The solid (dark blue), dot-dashed (blue), dashed (light blue) and dotted (lighter blue) lines refer to $r=1$, $r=1/2$, $r=1/3$ and $r=1/4$, respectively. }\label{fig2} \end{figure}

{III.3. \it Third model.} The next model we investigate is related to the subset of modified gravity with Lagrange multiplier defined by mimetic gravity \cite{15}. We reach those systems when taking the $\omega(\phi)$ in the form $\omega(\phi)=\phi,$
and consequently the potential  $ V(\phi)$ becomes constant. In this case the constraint of the model given in equation \eqref{constrain} has the form 
\be \label{constrain3}
g^{ab}\partial_{a}\phi\partial_{b}\phi=-1,
\ee
which is exactly the bound emerging from the metric factorization originally proposed within this setup \cite{16,17} and responsible for the decoupling of the conformal mode and the departure from general relativity. We remark here that our understanding of mimetic gravity is different from that presented in \cite{34}, and we refer to it only in the case where the constraint \eqref{constrain3} is satisfied. This choice implies that for any choice for $W(\phi)$ we have to deal with the equations 
\begin{equation}
\phi=y,\;\;\;I(\phi)=\frac{dW}{d\phi},\;\;\;U(\phi)=\dfrac{dW}{d\phi}-\frac{4}{3}W^{2}\,.
\end{equation}
Let us consider the case where $W$ is 
\be
W(\phi)=(3r/2)\,\tanh(\phi),
\ee
with $r$ being real parameter. With this choice for $W$ the Lagrange multiplier becomes $I(\phi)=(3r/2)\sech^2(\phi)$ and the warp factor gets the form 
\be
e^{2A}=\sech^{2r}(y).
\ee
It is now straightforward to obtain the energy density associated with this model; it has the form
\be
\rho(y)=\frac{3r}{2}\sech^{2r}(y)\left((2r+1)\sech^{2}(y)-2r)\right).
\ee
The warp factor and the energy density are depicted in Fig. \ref{fig2} for some values of $r$, and we see that they behave adequately, even though the scalar field diverges asymptotically.

We also notice from the behavior of the warp factor shown in Fig. \ref{fig2}, that it becomes thinner, as $r$ increases. This effect is similar to the effect unveiled before in \cite{31}, in the study on the possibility to make the brane compact, that is, to shrink the warp factor inside a compact region along the extra dimension. The effect here is similar, and we have checked that when $r$ increases to higher and higher values, the warp factor in fact shrinks to a smaller and smaller region around the center of the solution along the extra dimension. However, the Kretschmann scalar behaves inappropriately in the limit $r\to \infty$, so we cannot probe the compact limit of the brane in the present context. 
\begin{figure}[t!]
\begin{center}
\subfigure[$~$ Stability potential of the first model]{\includegraphics[width=0.8\linewidth]{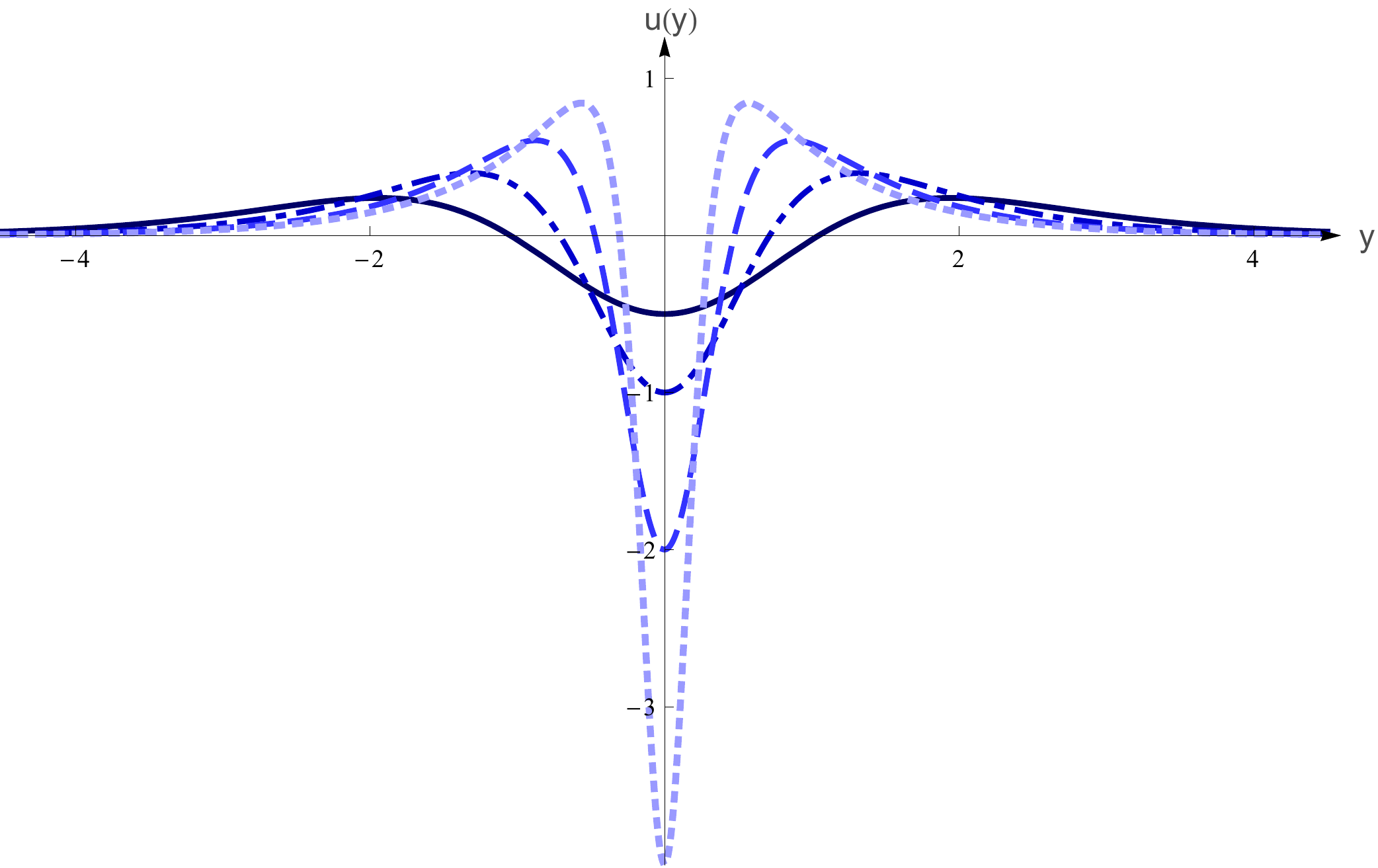}\label{fig4a}} \hspace{5mm}
 \subfigure[$~$ Stability potential of the second model]{\includegraphics[width=0.8\linewidth]{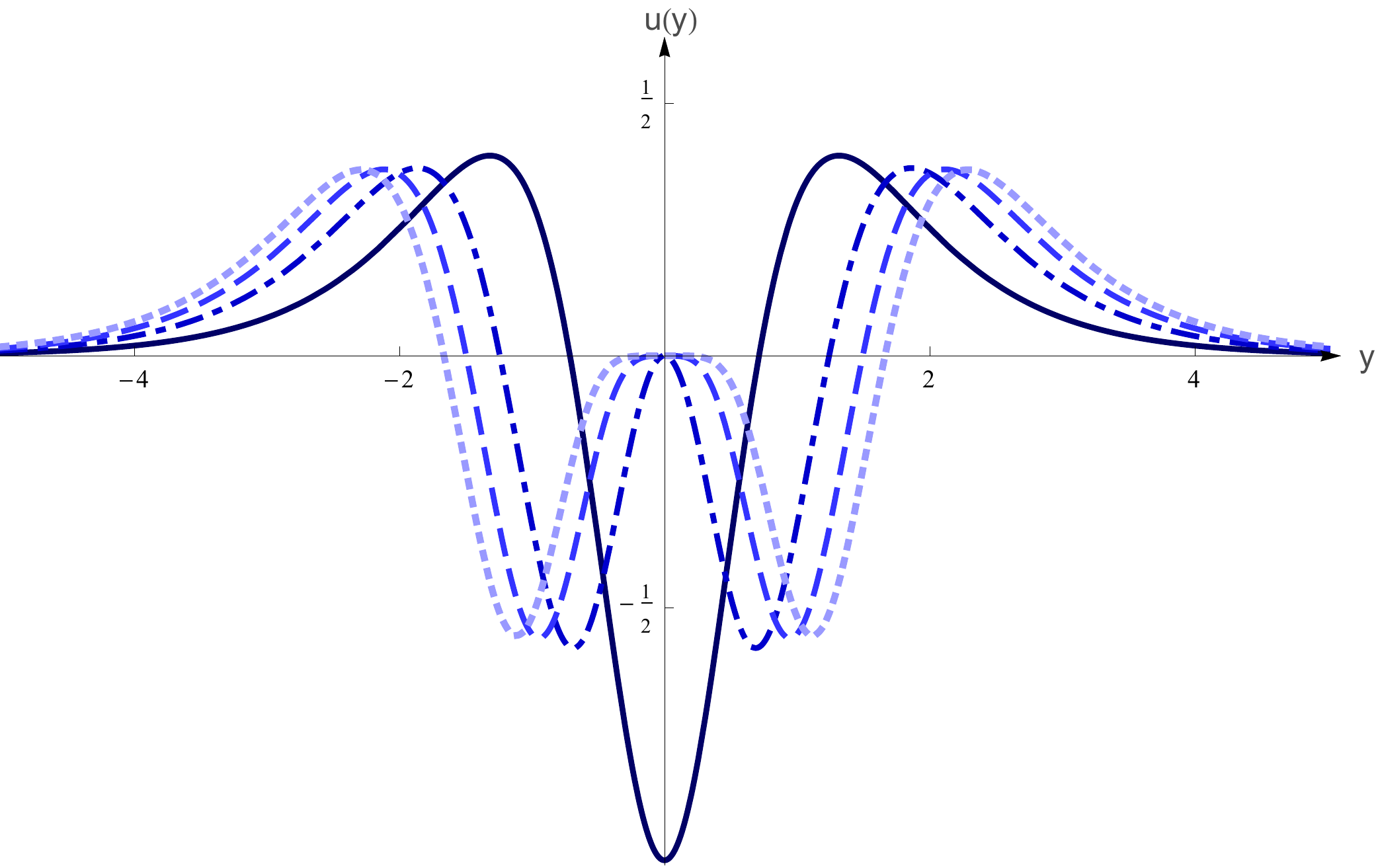}\label{fig4b}} \hspace{5mm}
\subfigure[$~$ Stability potential of the third model]{\includegraphics[width=0.8\linewidth]{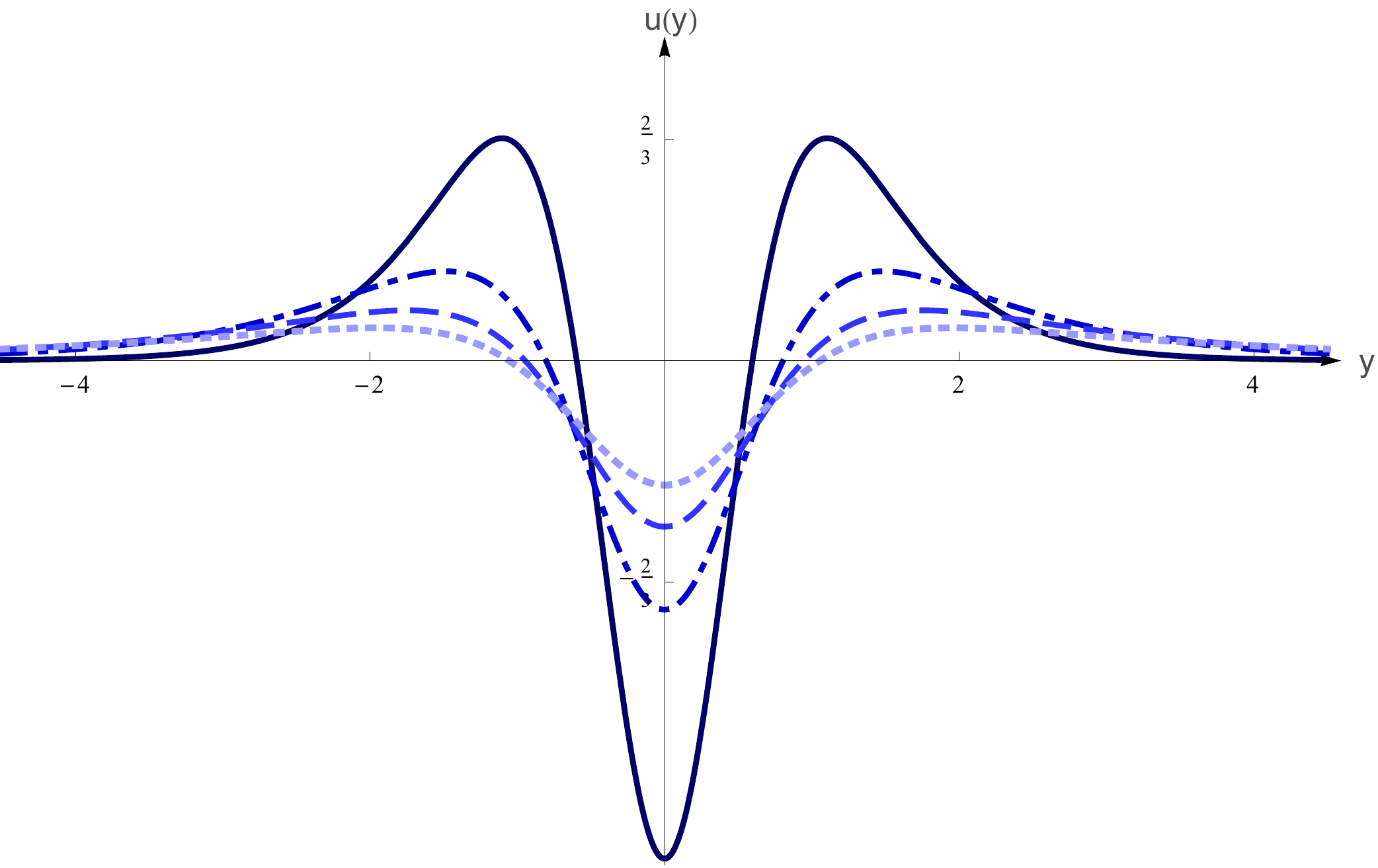}\label{fig4c}} 
\end{center}
\caption{Stability potential of the three models investigated in Sec. III. They are depicted using the same conventions of their corresponding
Figs. \ref{fig0}, \ref{fig1} and \ref{fig2}.}\label{fig4}
 \end{figure}

{IV. \it Stability.} We now proceed to investigate the stability of the gravitational sector of the models. We follow the investigation described in \cite{20} and first make a coordinate redefinition for the extra dimension given by $ dy=e^{A(z)}dz$,  so that now the background geometry is conformally flat, i.e., $ \tilde{g}_{ab}=e^{2A(z)}\eta_{ab}$. We then take small fluctuations around the metric by making the change
$\eta_{ab}\to\eta_{ab}+h_{ab}(x,z)$, with $h_{55}=0$. These fluctuations describe the gravitational behavior along the extra dimension around the brane. In this way, the perturbed metric becomes
\be \label{2.1}
ds^{2}=e^{2A(z)}(\eta_{ab}+h_{ab})dx^{a}dx^{b}\,.
\ee
The linear correction to the Einstein tensor in the transverse traceless gauge ($\partial_\mu h^{\mu\nu}=0$ and $h_\mu^\mu=0$) due to the presence of gravitational fluctuations is 
\be \label{2.22}
\delta G_{\mu\nu}=-\frac{1}{2}\partial_{c}\partial^{c}h_{\mu\nu}+\frac{3}{2}{\dot A}{\dot h}_{\mu\nu}-3h_{\mu\nu}({\ddot A}+{\dot A}^{2})\,,
\ee
and $\delta G_{55}=0$ (since $h_{55}=0$), where dot represents derivative with respect to the new coordinate $z$. In order to find the equation describing the gravitational modes in the metric background \eqref{2.1} we have to solve the general equation $\delta G_{\mu\nu}=2\delta T_{\mu\nu}$. One can prove that the linear correction to the energy-momentum tensor is $ 2\delta T_{\mu\nu}=-3h_{\mu\nu}({\ddot A}+{\dot A}^{2})$ and then the equation becomes 
\be \label{2.3}
\partial_{c}\partial^{c}h_{\mu\nu}-3{\dot A}{\dot h}_{\mu\nu}=0.
\ee
In order to clarify how the above equation relates to the stability of the system under small fluctuations, we take $ h_{\mu\nu}=e^{-imx}e^{3A/2}H_{\mu\nu} $, so we are led to the Schr\"odinger-like stability equation
\be \label{2.4}
\left(-\partial^{2}_{z}+u(z)\right) H_{\mu\nu}=m^{2}H_{\mu\nu}\,,
\ee
where the stability potential is given by
\be \label{stability1}
u(z)=\frac{3}{2}{\ddot A}+\frac{9}{4}{\dot A}^{2}\,.
\ee
The stability potentials of the models presented in the previous section are depicted in Fig. \ref{fig4}, where one can realise how they react to the variations of the parameters of the models. We notice that in Fig. \ref{fig4}, the warp factors are displayed in terms of $y$, so they were depicted using the stability potential
\be  
u(y)=\frac34 e^{2 A}(2 A''+ 5 A'^2).
\ee
Fig. \ref{fig4a} shows a standard behavior, but in Fig. \ref{fig4b} the stability potential of the second model has local maxima in the central region of the potential for $n \neq0$, which results from the splitting of the branes in that case. In Fig. \ref{fig4c} the parameter $r$ of the third model controls the thickness of the potential and its depth, but do not have an internal structure.

The stability equation \eqref{2.4} can be factorized as
\be\label{factesteq}
\left(\partial_{z}+\frac{3}{2}{\dot A}\right)\left(-\partial_{z}+\frac{3}{2}{\dot A}\right) H_{\mu\nu}=m^{2}H_{\mu\nu}\,.
\ee
Thus, if we define $Q=-\partial_z+(3/2) {\dot A},$ it is possible to rewrite \eqref{factesteq} in the form $
Q^\dag Q H_{\mu\nu}=m^2 H_{\mu\nu}.$ In this way, the operator $Q^\dag Q$ is nonnegative and so $m^2\geq 0$, forbidding the presence of states with negative eigenvalue. The gravitational sector of the model is then linearly stable. In particular, the graviton zero mode is proportional to $e^{3A(z)/2}$. Gravity localization then requires the zero mode to be normalisable, i.e., the integral of $\exp(3A(z))$ over $z\in(-\infty,\infty)$ should be finite. This condition implies that the warp function $A(z)$ has to behave properly to ensure finiteness of the above integral.

In the case of the thin brane which is obtained with $\omega(\phi)=\phi$, or with $W(\phi)=\phi$, as discussed at Sec. III, the  stability potential is given by  
\be \label{stability3}
u(z)=\frac{15}{4}\frac{\kappa^{2}}{(\kappa|z|+1)^{2}}-3\kappa\delta(z)\,,
\ee
and its stability analysis was already performed in \cite{18}.

{V. \it Conclusion.} In this work we developed a first order formalism for thick branes in the context of modified gravity with Lagrange multiplier. The procedure was implemented with two auxiliary functions $\omega(\phi)$ and $W(\phi)$, which allowed us to solve the equations of motion with first order differential equations. Interestingly, the use of such formalism brings us new possibilities, illustrated with distinct systems. The first model presented a standard scenario, controlled by the parameter $\alpha$, and we noticed that the limit of a very large $\alpha$ provided another way to get back to the thin brane scenario firstly discussed in \cite{18}. The second model is also of interest, since it leads us to the case of brane splitting, which is characterized by the peculiar profile of energy density when $ n=1,2,3,$ etc. The third model is also interesting since it leads to the case of thick mimetic branes, with the scalar field giving rise to a solution that diverges asymptotically, but capable of producing a thick brane with the standard profile. This model, in particular, engenders the effect of allowing the brane to shrink inside a compact region around the center of the solution along the extra dimension. We have also examined linear stability of the gravity sector of the models, showing that the braneworld scenarios are stable against small fluctuations of the metric.

The first order procedure developed in this work shows not only that the addition of Lagrange multipliers is a useful tool to construct new thick branes models, but also that it unveils a route to describe analytical solutions, which are welcome since they allow a clearer view of the problem. In particular, it can perhaps be extended to other scenarios, as the one with two or more scalar fields. Given the form of the potential $U(\phi)$ that appears in \eqref{potU}, another possibility of investigation concerns the construction of asymmetric branes, which can be implemented with the methodology recently used in \cite{47} to explore the problem. The first order procedure may also be useful to examine the domain-wall/brane-cosmology correspondence \cite{37} in the new scenario of modified gravity with Lagrange multiplier. We hope that the above results will foster new investigations in this subject. 

{\it Acknowledgments.} D.B. and D.A.F would like to thank CAPES, CNPq and Para\'\i ba State Research Foundation (FAPESQ/PB, Grant 0015/2019) for partial financial support.

\end{document}